\documentclass[a4,11pt]{article}

\usepackage[dvips]{graphicx}

\title{Observation of ER UMa Stars}

\author{Taichi KATO, Daisaku NOGAMI, Hajime BABA, \\ Seiji MASUDA, 
Katsura MATSUMOTO \\
{\it Department of Astronomy, Kyoto University, } \\
{\it Sakyo-ku, Kyoto 606-8502, Japan, } \\
{\it (tkato,nogami,baba,masuda,katsura)@kusastro.kyoto-u.ac.jp} \\
Chatief KUNJAYA \\
{\it Department of Astronomy, Institut Teknologi Bandung, } \\
{\it Ganesha 10, Bandung 40132, Indonesia, } \\
{\it KUNJAYA@sirius.as.itb.ac.id}
}

\date{}

\begin{document}

\maketitle

\section*{Abstract}

   ER UMa stars are a recently recognized small subgroup of SU UMa-type
dwarf novae, which are characterized by the extremely high outburst
frequency and short (19--48 d) supercycles.  From the current thermal-tidal
disk instability scheme, they are considered to be high mass-transfer
SU UMa-type dwarf novae, and comprise a link to {\it permanent superhumpers}
below the period gap.  They do not only provide an opportunity to test
the applicability of thermal-tidal instability model but also pose
problems on the origin of high mass-transfer in short orbital-period
cataclysmic variables.  A historical review of this subgroup and recent
topics of ER UMa stars, the unique pattern of superhump evolution
and the ``helium ER UMa analog" (CR Boo), are also discussed.

\section{Introduction}

   Before the discovery of ER UMa stars (or RZ LMi stars), there had been
a well-known standard (or {\it classical}) picture of SU UMa-type dwarf
novae.  SU UMa-type dwarf novae, a subclass of dwarf novae, show two
types of outbursts --- normal outbursts and superoutbursts --- which are
now explained by the fruitful combination of thermal and tidal instabilities
(Osaki 1989).  During superoutbursts of SU UMa-type dwarf novae, superhumps
appear which are a result of the tidal instability.  In the canonical
view of SU UMa-type dwarf novae, the mass-transfer from the secondary
is governed by the angular momentum loss by the gravitational wave radiation
(GWR), which is a strong function of the secondary mass, or the orbital
period.  SU UMa-type dwarf novae were thus considered to be a
{\it one-parameter system}, whose system parameters are a strong function
of the orbital period.

   There have been classically known exceptions to this view.
One is what is called ``WZ Sge stars", which show extremely
low outburst frequencies and the predominance of superoutbursts.
There have been arguements whether WZ Sge stars can be understood as
a natural extension of SU UMa-type dwarf novae to the lowest mass-transfer
end, or whether there are a need for other relevant mechanisms or physics.
The other is the presence of novalike (NL) stars (and classical novae)
{\it in the period gap} or {\it below the period gap}, suggesting the
presence of systems having mass-transfer rates higher than the
canonical picture.  The unambiguous interpretation of superhump-like
modulation in the NL stars was difficult because they can be also
interpreted as the intermediate polar effect.  The discovery of ER UMa
star has drastically changed the canonical picture of SU UMa-type
uniformity, and provided a missing, and smooth link between (classical)
SU UMa-type dwarf novae and NL stars below the period gap.

\begin{figure}[t] 
\begin{center}
  \includegraphics[width=12cm,clip]{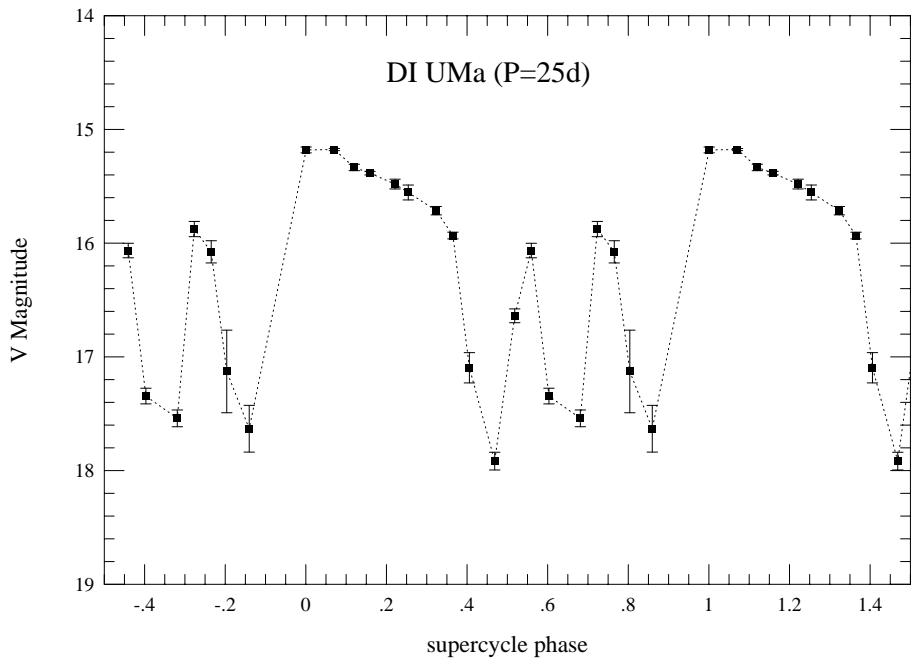}
  \caption{Representative light curve of an ER UMa star (DI UMa)} 
\end{center}
\end{figure}

\section{History of ER UMa stars}

   Back to ER UMa itself, ER UMa (PG~0943+521) was discovered as an
ultraviolet-excess object (Green et al. 1986).  The lack of variability
information classified it as an inconspicuous novalike cataclysmic binary.
However, by several groups differently and independently motivated, the
unique variability of this object began revealing itself in early the 1990's.
Among them, M. Iida (VSOLJ) in 1992 was the first to visually discover
dwarf-nova outbursts in this object (Iida 1994).
His subsequent observations led to the present variable star designation
(ER UMa).  Follow-up collaborations by amateur and professional observers
tentatively classified the object as an Z Cam-type dwarf nova as judged
from the presence of ``standstill"-like plateau stages.
Another was by a systematic follow-up study of the only complete
(magnitude-limited) sample of cataclysmic variables (CVs):
the PG survey.  Misselt and Shafter (1995) detected the photometric period of
1.6 hour in ER UMa, but its origin was not fully pursuited at the time
of initial discovery.  The third was the product of the robotic telescope
(RoboScope), which enabled long-term automated observations of NLs and old
novae, particularly motivated to search for an evidence of temporary
changes of mass-transfer in these systems.  Robertson et al. (1995)
discovered a peculiar outburst pattern in ER UMa, V1159 Ori, RZ LMi and
V446 Her.  The outburst pattern is characterized by the stability of
outburst cycle length, and nearly strictly repeating bright states.
They initially interpreted this phenomenon by the periodic mass-transfer
burst from the secondary (Honeycutt et al. 1994, a talk given at the
Abano-Terme CV Conference): the long-lasting bright state corresponds to
the novalike state, when the disk is thermally stable, while
the frequently outbursting state corresponds to the state of reduced
mass-transfer, giving rise to the dwarf nova-type disk instability.

\begin{figure}[t]
\begin{center}
   \includegraphics[width=12cm,clip]{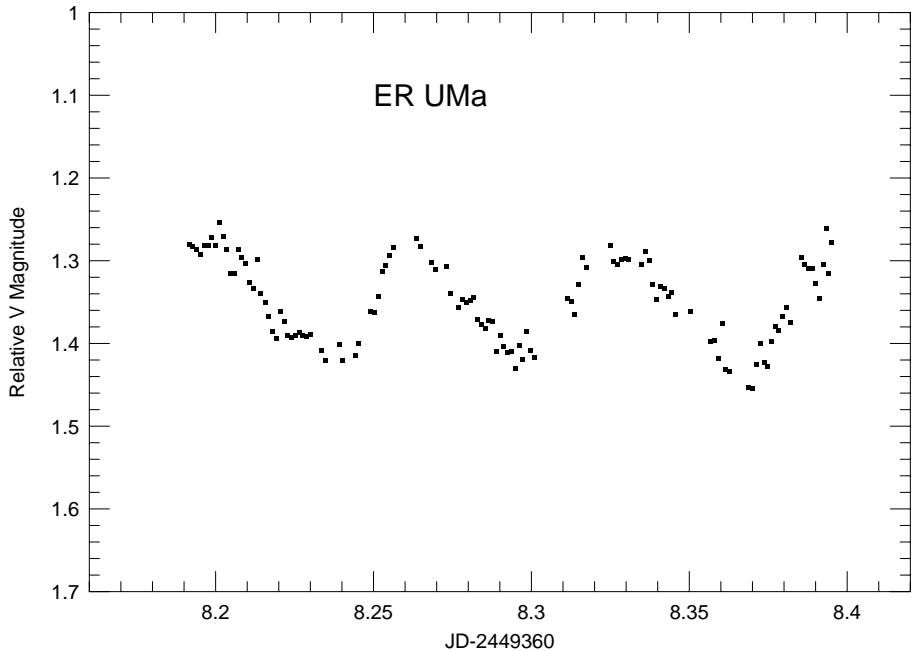}
   \caption{Superhumps in ER UMa}
\end{center}
\end{figure}

   The object finally revealed its nature in early 1994, led by the
detection of a bright outburst by the amateur observer, G. Poyner.
A ``snapshot" CCD time-series observation at Ouda Station showed
some hint of modulations; the signal grew night by night, and fully
grown superhump finally emerged (Fig. 2).  Compilation of visual and CCD
observations, via pre-VSNET amateur-professional collaborations, revealed
the 43-d recurrence of superoutbursts (Kato and Kunjaya 1995).
ER UMa was thus confirmed to be an SU UMa-type dwarf nova, and dramatically
broke the shortest record of supercycle length $T_s$ (the short known
at that time was $T_s \sim$130 d for YZ Cnc).
The subsequently discovery of the ``ER UMa-twin", V1159 Ori
(Nogami et al. 1995a, 1995b; this object was independently studied by
Patterson et al. 1995),
and the detection of superhumps in RZ LMi (Robertson et al. 1995;
Nogami et al. 1995b) quickly led to a concept of small subgroup of
SU UMa-type dwarf novae having the shortest supercycle lengths (Fig. 3
and Table 1).  Up to now, another member, DI UMa, has been added
(Kato et al. 1996a).

\section{Questions arisen by ER UMa stars}

   The discovery of ER UMa stars soon invoked several questions previously
disregarded in the canonical context of SU UMa-type dwarf novae.
The first question was whether such short supercycles (48 to 19 d) can be
reproduced by the standard disk instability model for SU UMa-type dwarf
novae (Osaki 1989).  Osaki (1995a) could reproduce the light curve of
ER UMa by increasing the mass-transfer rate by a factor of $\sim$10.
In his model, supercycles longer than $\sim$40 d could be reproduced;
Osaki (1995a) further revealed that as the mass-transfer rate increases
beyond this point, the durations of superoutbursts and supercycles lengthen,
and finally systems reach a thermally stable, but tidally unstable state.
Osaki (1995a) considered this state corresponds to ``permanent superhumpers".
Osaki (1995b) also showed the shortest case of 19-d supercycle (RZ LMi)
can be reproduced by assuming the reduced tidal torque, whose physical
origin has not yet been well understood.

\begin{figure}[t]
\begin{center}
   \includegraphics[width=12cm,clip]{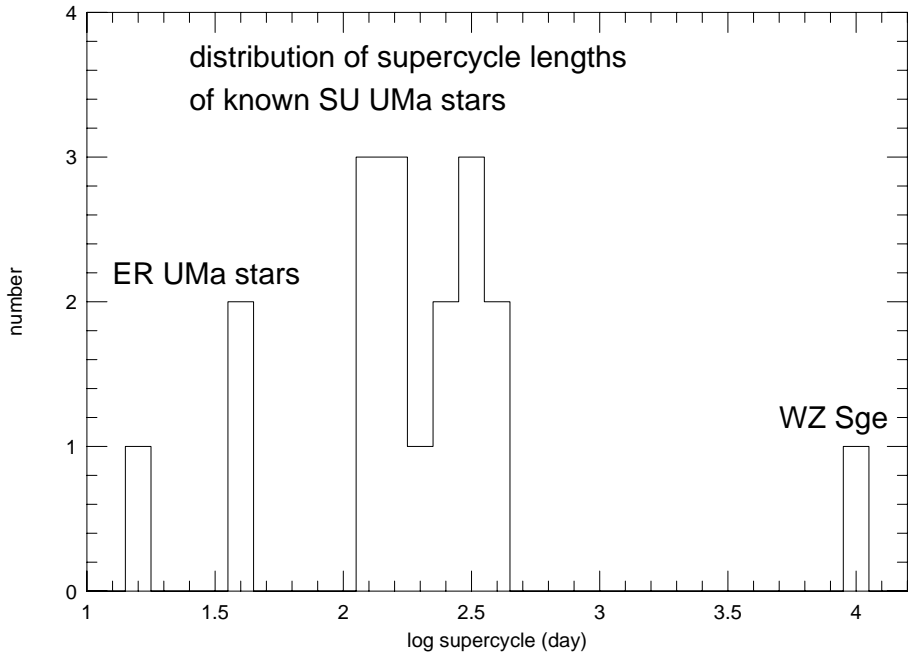}
   \caption{Distribution of supercycle among SU UMa stars.  ER UMa stars
   occupy the separated region below $T_s \sim$50 d.
   }
   \caption{Superhumps in ER UMa}
\end{center}
\end{figure}

   The second is the origin of high mass-transfer rates.  In the canonical
view, the evolution and the mass-transfer below the period gap are
primarily governed by the gravitational wave radiation.  The supercycle
length, which, in the scheme of thermal-tidal disk instability, is roughly
inversely proportional to the mass-transfer rate, is expected to be a
strong function of the orbital period.  ER UMa stars apparently break
this view (Fig. 4).  Several possibilities have been raised, including the
angular momentum loss other than GWR (e.g. weakly persistent magnetic
breaking) and the temporary increase of mass-transfer rates triggered by
nova explosions.  It is not yet clear which mechanism is responsible,
and the question remains why ER UMa stars are so concentrated near
the CV period minimum.

\begin{figure}[t]
\begin{center}
   \includegraphics[width=12cm,clip]{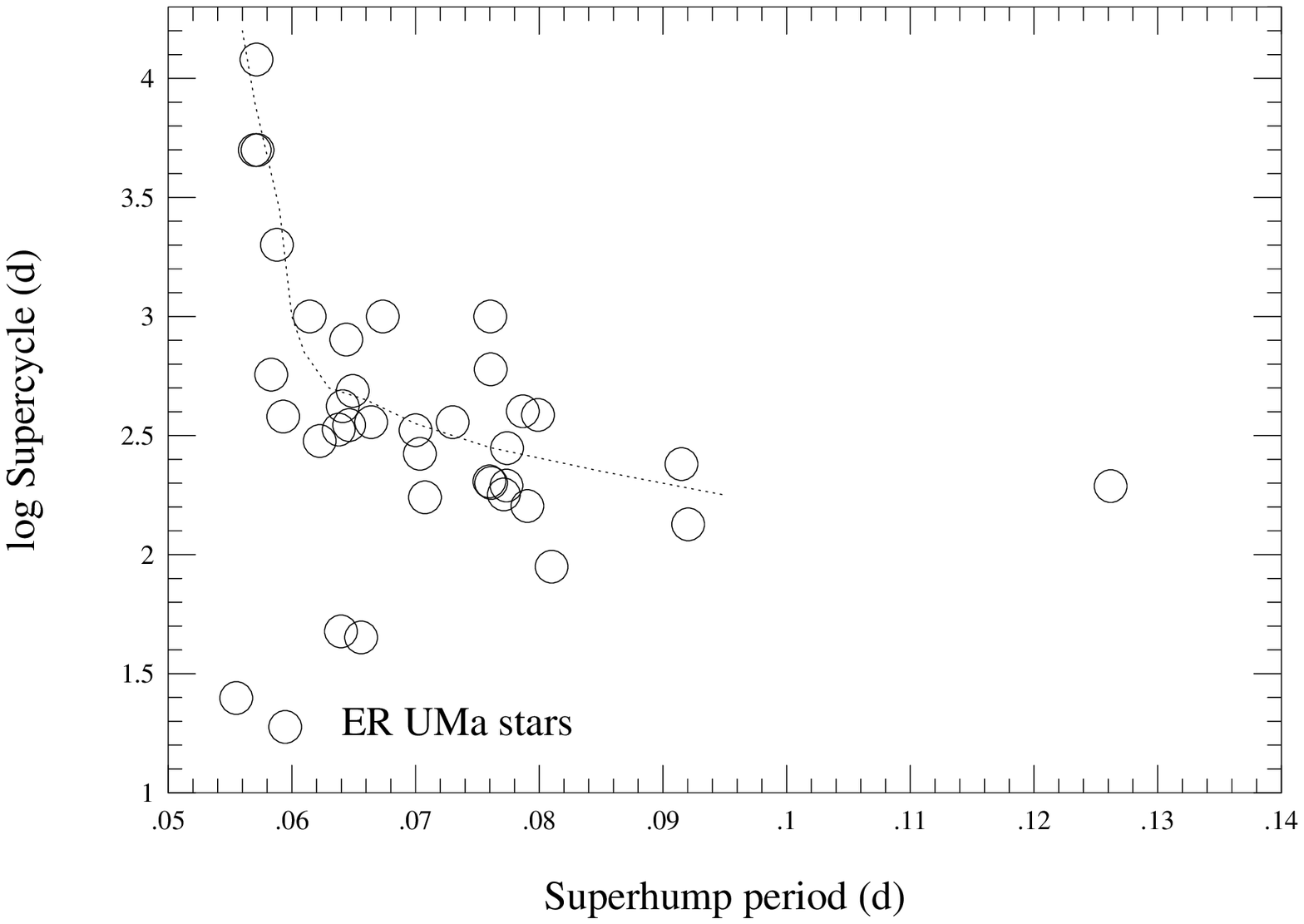}
   \caption{Distribution of supercycle among SU UMa stars.  ER UMa stars
   occupy the separated region below $T_s \sim$50 d.
   }
   \caption{Superhump period versus supercycle length.  Usual SU UMa stars
   and WZ Sge stars are on the dashed line, while ER UMa stars occupy
   a small separated region.
   }
\end{center}
\end{figure}

   The third is the (space) number density problem.  The number of known
ER UMa stars is four out of $\sim$40 known SU UMa-type dwarf novae.
The evolutionary time scale being inversely proportional to the
mass-transfer rate, the probability of encountering such systems is
$1 \over 10$ of usual SU UMa-type dwarf novae, if ER UMa stars are
truly high mass-transfer systems.  The observed numbers (although
statistically incomplete) suggests either 1) if ER UMa stars are
evolutionally distinct populations from usual SU UMa-type dwarf novae,
the source (progenitor) of such systems should be as numerous as
usual SU UMa-type stars, or 2) if ER UMa stars and usual SU UMa-type
stars are interchanging states of the same population, systems must
accrete half of mass during the ER UMa state.  The discussion on
the impact on the CV evolution below the gap should require
further observational constraints, e.g. independent estimates of
mass-transfer rates in ER UMa stars, unbiased surveys of dwarf novae
and searches for secular changes as expected from nova-dwarf nova
alternations.

\begin{figure}[t]
\begin{center}
   \includegraphics[width=12cm,clip]{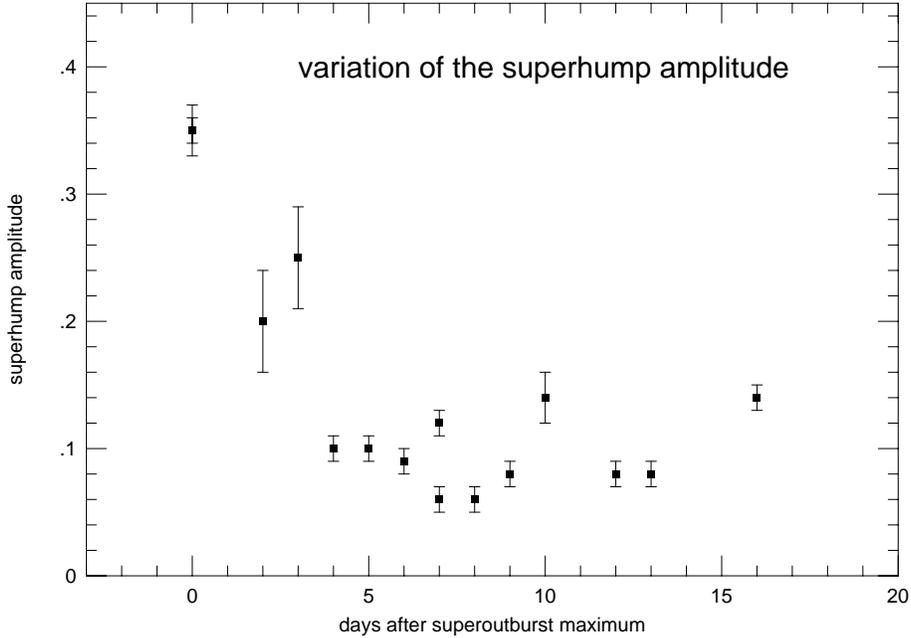}
   \caption{Variation of superhump amplitudes in ER UMa.  Large-amplitude
   superhumps appear during the earliest stage of superoutburst, and decay
   rapidly.
   }
\end{center}
\end{figure}

\section{Large-amplitude superhumps during early superoutburst}

   ER UMa stars in common additionally have other peculiar features
in contrast to usual SU UMa-type dwarf nova.  Among them is the appearance
of large-amplitude early stage superhumps (Kato et al. 1996b).
These superhumps are seen in all known ER UMa stars, and quickly develop
during the rise to superoutbursts.  They quickly decay in a few days, and
usually-looking superhumps appear (this growth was observed at the discovery
moment of ER UMa) in a week (Fig. 5), but the phase is reversed to the
initial superhumps.  Large-amplitude early (super)humps and the later growth
of usual superhumps are also observed in WZ Sge-type stars (e.g. AL Com,
Kato et al. 1996c), but the main difference is that the superhump period
remain basically unchanged through the superhump evolution in ER UMa stars.

\begin{figure}[t]
\begin{center}
   \includegraphics[width=12cm,clip]{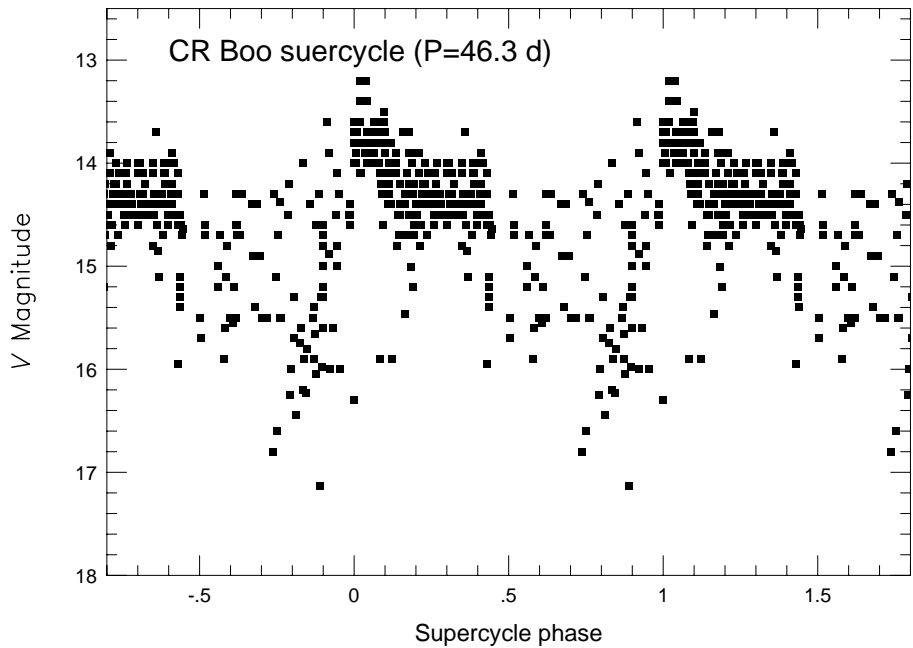}
   \caption{Light curve of the ``helium ER UMa star", CR Boo folded by
   the supercycle 46.3 d.  The alternation of long superoutburst and
   short outbursts is remarkably similar to that of ER UMa.
   }
\end{center}
\end{figure}

\section{Helium ER UMa stars?}

  Double-degenerate helium CVs, also called AM CVn stars, are considered
as helium counterparts to hydrogen-rich (usual) CVs.  Among them, AM CVn
and HP Lib correspond to superhumping novalike variables (permanent
superhumpers), CR Boo, V803 Cen and CP Eri are SU UMa-type analogs, and
GP Com may be a helium counterpart to WZ Sge stars.  We have recently shown
(Kato et al. 1998) CR Boo shows 46.3-d supercycle and its behavior closely
analogous to ER UMa itself (Fig. 6).  V803 Cen may be a similar object having
a likely supercycle of $\sim$60 d, but also seems to show different
(ER UMa-like and quasiperiodically rapidly outbursting) states.

\begin{table}
    \caption{Known (hydrogen) ER UMa stars} 
\vspace{.5pc}
\begin{center}
\begin{tabular}{l|cccc} \hline
            & ER UMa      & V1159 Ori  & RZ LMi      & DI UMa  \\
\hline
other name  & PG~0943+521 & NSV~02011  & PG~0948+344 & NSV~04407 \\
maximum (V) & 12.4        & 12.8       & 14.2        & 15.1     \\
minimum (V) & 15.2        & 15.4       & 17.0        & 18.0     \\
P$_{\rm SH}$ (d) & 0.06573 & 0.0641    & 0.05946     & 0.0555   \\
T$_s^*$ (d)  & 43         & 48         & 19          & 25       \\
T$_n^\dag$ (d) & 4.4      & 4.0        & 3.8         & 4        \\
\hline
\end{tabular}
\\
\end{center}
$^*$ supercycle length \\
$^\dag$ recurrence time of normal outbursts \\
For more information of the individual light curves, see
http://www.kusastro.kyoto-u.ac.jp/vsnet/LCs/index/index.html. \\
\end{table}

\newpage

\section{References}

\noindent

1. Green R. F., Schmidt M., Liebert J.\ 1986, ApJS 61, 305

2. Iida M.\ 1994, VSOLJ Variable Star Bulletin 19, 2

3. Kato T., Kunjaya, C.\ 1995, PASJ 47, 163

4. Kato T., Nogami D., Baba H.\ 1996a PASJ 48, L93

5. Kato T., Nogami D., Masuda S.\ 1996b PASJ 48, 5

6. Kato T., Nogami D., D., Baba H., Matsumoto K., Arimoto J., Tanabe K.,
      Ishikawa K.\ 1996c PASJ 48, L21

7. Kato T., Nogami D., Baba H., Hanson G., Poyner G.\ 1998 MNRAS submitted

8. Misselt K. A., Shafter A. W.\ 1995, AJ 109, 1757

9. Nogami D., Kato T., Masuda S., Hirata R.\ 1995a, IBVS No.4155

10. Nogami D., Kato T., Masuda S., Hirata R., Matsumoto K., Tanabe K.,
     Yokoo T.\ 1995b, PASJ 47, 897

11. Osaki Y.\ 1989, PASJ 41, 1005

12. Osaki Y.\ 1995a, PASJ 47, L11

13. Osaki Y.\ 1995b, PASJ 47, L25

14. Patterson J., Jablonski F., Koen C., O'Donoghue D., Skillman D. R.\
     1995, PASP 107, 1183

15. Robertson J. W., Honeycutt R. K., Turner G. W.\ 1995, PASP 107, 443

\section*{{\it Note on the online version}}

{\it This is a LaTeX typeset of the proceeding paper Disk Instabilities 
in Close Binary Systems. 25 Years of the Disk-Instability Model. 
Proceedings of the Disk-Instability Workshop held on 
27-30 October, 1998, at Hotel Brighton City, Yamashina, Kyoto, Japan. 
Edited by S. Mineshige and J. C. Wheeler. Frontiers Science Series 
No. 26. Universal Academy Press, Inc., 1999., p.45.
(1999dicb.conf...45K).  Since the original style file was an old one,
I used a standard LaTeX class to reproduce it.  Although the layout is
different from the one in the print, the text and figures are
the original ones.}

\end{document}